\documentclass[usenatbib,twocolumn,useAMS]{mn2e}
\usepackage{graphicx}
\usepackage{latexsym}
\usepackage{dcolumn}% Align table columns on decimal point
\usepackage{bm}% bold math
\input{epsf}

\newcommand{\bfig}{\noindent\begin{minipage}{3.48in}}
\newcommand{\efig}{\bigskip\end{minipage}}
\newcommand{\be}{\begin{equation}}
\newcommand{\ee}{\end{equation}}
\newcommand{\ba}{\begin{eqnarray}}
\newcommand{\ea}{\end{eqnarray}}
\newcommand{\bi}{\begin{itemize}}
\newcommand{\ei}{\end{itemize}}
\def\mnras{MNRAS}
\def\apj{ApJ}
\def\apjl{ApJL}
\def\apjs{ApJS}

\def\araa{Annu. Rev. Astron. Astrophys.}

\def\aap{Astron. \& Astrophys.}

\title[Cosmic magnification and 21cm emitting galaxies]{Precision
  measurement of cosmic magnification from  21 cm 
  emitting galaxies}
\author[Pengjie Zhang \& Ue-Li Pen]
{Pengjie Zhang$^1$\thanks{E-mail:zhangpj@fnal.gov},
  Ue-Li Pen$^2$\thanks{E-mail:pen@cita.utoronto.ca}\\
$1$NASA/Fermilab Astrophysics Group,
Fermi National Accelerator Laboratory,
Box 500,
Batavia, IL 60510-050\\
$2$Canadian Institute for Theoretical Astrophysics, University of
Toronto, Toronto, Canada, M5S 3H8}

\begin{document}
\maketitle
\begin{abstract}
We show how precision lensing measurements can be obtained through the lensing
magnification effect in high redshift 21cm emission from galaxies.  Normally,
cosmic magnification measurements have been seriously complicated by galaxy
clustering.  With precise redshifts obtained from 21cm emission line
wavelength, one can correlate galaxies at different source 
planes, or exclude close pairs to eliminate such contaminations.

We provide forecasts for future surveys, specifically the SKA and CLAR.  SKA
can achieve percent precision on the dark matter power spectrum and the galaxy
dark matter cross correlation power spectrum, while CLAR can measure
an accurate cross correlation
power spectrum.  The neutral hydrogen fraction was most likely significantly
higher at high redshifts, which improves the number of observed
galaxies  significantly, such
that also CLAR can measure the dark matter lensing power spectrum. SKA
can also allow precise measurement of lensing bispectrum.  
\end{abstract}

\section{Introduction}
Gravitational lensing measures the distortion of light by
gravity originating from  inhomogeneous distribution of matter. The physics of
weak gravitational lensing is clean, perhaps comparable to the primary
CMB. To the first order 
approximation, it  involves only general relativity and collision-less
dark matter dynamics. Gas physics only enters at small scales
\citep{Zhan04,White04} and can be nulled by throwing away  small
scale  lensing information
\citep{Huterer05}. Though the prediction of weak lensing statistics is
complicated  by nonlinear evolution of  matter density fluctuation, high
resolution, large box size N-body simulations are able to measure
these  statistics to high accuracy (e.g. \citet{Vale03,Merz05}). Thus,  
weak gravitational 
lensing is one of  the most powerful and robust tools  
to constrain cosmology and study the large scale structure of the
universe.

Weak gravitational lensing  has been detected through cosmic shear (see
\citet{Refregier03} for a recent review). 
%In the last several years,
%many groups have reported successful detections of cosmic shear 2-point
%statistics such as power spectrum and variance in various optical surveys
%\citep{Bacon00,Kaiser00,vanWaerbeke00,Wittman00,Maoli01,Rhodes01,vanWaerbeke01, 
%Hoekstra02a,Hoekstra02b,Refregier02,Bacon03,Brown03,Hamana03,Jarvis03}.
%Lensing systematics can be quantified by  the E-B decomposition
%\citep{Kaiser92,Stebbins96,Crittenden02,Pen02} and has 
%been shown to be under control. Higher order statistics such as cosmic
%shear convergence skewness has also been robustly detected
%\citep{Bernardeau02,Pen03,Jarvis04}.  Cosmic shear has 
%also been detected in radio 
%surveys  \citep{Chang04} and in galaxy-galaxy lensing
%\citep{Hirata04,Sheldon04}.   
Ongoing and upcoming large scale galaxy 
surveys such as
CFHTLS\footnote{http://www.cfht.hawaii.edu/Science/CFHLS/},
DES\footnote{http://www.darkenergysurvey.org/},
LSST\footnote{http://www.lsst.org/},
Pan-STARRS\footnote{http://pan-starrs.ifa.hawaii.edu/public/science/} and 
SNAP\footnote{http://snap.lbl.gov/}  will reduce the statistical
errors of the cosmic shear measurement to $\la 1\%$ level. At that
stage, systematic errors such as point spread function,
galaxy intrinsic  alignment, seeing and extinction 
\citep{Hoekstra04,Jarvis04,Vale04,vanWaerbeke04} and errors in  galaxy
redshift measurement will be  ultimate limiting factors.  Given these possible
systematics, it is worthwhile to search for independent method to
measure the gravitational lensing, for consistency check and, for
likely better statistical and systematic errors. In this paper, we
will show that  the cosmic magnification of 21cm emitting galaxies is
a competitive candidate.

Cosmic magnification is the lensing induced changes in galaxy number
density. 
%Besides distorting galaxy images, gravitational lensing also changes
%the background galaxy number density. This effect is two-fold. It increases
%(decreases) the area 
%of a given patch on the 
%sky and thus tends to decrease (increase) galaxy number
%density. On the other hand, it increases (decreases) galaxy
%flux. Since galaxy surveys are 
%flux limited, this results in more (less) {\it observed} galaxies in
%a given patch of the sky. The combined effect is either enhancement or
%suppression 
%of {\it observed} galaxy density, depending on the slope of galaxy
%luminosity function at the observation flux limit. This magnification
%effect is called cosmic magnification. 
It 
introduces extra correlations in 
galaxy clustering \citep{Kaiser92, Villumsen96, Moessner98a,Jain03} and
correlates galaxies (quasars) at widely separate redshifts \citep{Moessner98b}. Cosmic magnification contains as
much information of cosmology and matter clustering as cosmic
shear. It has been robustly detected in quasar-galaxy lensing
(\citet{Scranton05} and references therein). 
%\citep{Tyson86,Fugmann88,Fugmann90,Hintzen91,Thomas95,Bartelmann93, 
%Bartelmann94,Benitez95,Benitez97,Bartsch97,Norman99,Norman01,Menard03b,
%Myers03,Scranton05}.  Though early detections are often controversial,
%recent detection by \citet{Scranton05} shows clear dependence of signal
%on the shape of quasar luminosity function, as expected to be an
%unique signature of the cosmic magnification signal. 

The measurement of cosmic magnification does not require the accurate
determination of  
galaxy shapes and is thus free of many systematics, such as point
spread function and galaxy intrinsic alignment,  entangled in the
cosmic shear measurement.  But it 
suffers from several obstacles. (1) It
suffers stronger shot noise than 
cosmic shear measurement. For cosmic shear, shot noise comes from
the intrinsic ellipticity of galaxies, which has dispersion
$\langle \epsilon^2\rangle^{1/2}\simeq 0.3$. The shot noise power
spectrum is proportional to $\langle 
\epsilon^2\rangle/N_g\sim 0.1/N_g$, where $N_g$ is the total number of
galaxies. For cosmic magnification, shot 
noise comes from the Poisson fluctuation of galaxy counts and
the shot noise power spectrum scales as $1/N_g$ and is thus an order of
magnitude larger than that in cosmic shear. (2) The signal of cosmic
magnification is generally much smaller than intrinsic clustering of
galaxies. Without precise redshift measurement of galaxies or quasars,
the only method to remove intrinsic clustering of
galaxies is to measure the cross correlation of foreground galaxies
and background galaxies (or quasars). But its detection 
is severely limited  by the number of high redshift galaxies, which 
are difficult to detect in optical band, and quasars, which are rare,
comparing to galaxy abundance.  Early measurements of cosmic
magnification were often controversial. Even with relatively large
sample of quasars and foreground galaxies from 2dF and SDSS, the
measurement of cosmic magnification is still at its infancy and is
only confirmed to  $\leq 8\sigma$ confidence level.

Besides the observational difficulties, the theoretical prediction of cross
correlation strength is complicated by foreground galaxy bias, which is
hard to predict from first principles. Without precise understanding
of the galaxy bias, the power of cosmic magnification to constrain
cosmology and matter distribution is severely
limited, compared to cosmic shear.

 Cosmic magnification measured from  21cm emitting galaxies is free of 
many obstacles entangled in the cosmic shear measurement and
quasar-galaxy cosmic magnification measurement. 
 As we will show in \S \ref{sec:nz}, upcoming radio surveys such as
 Square Kilometer Array (SKA)\footnote{http://www.skatelescope.org/}
 can find $\sim 10^8$-$10^9$ HI-rich galaxies in  
total and $\sim 10^7$ galaxies at $z>2$ through the neutral
 hydrogen 21cm hyperfine transition line. With this large sample of
galaxies, the measurement of cosmic magnification will enter the
precision era, due to following reasons.  (1) Radio observations
 are free of 
extinction. Since dust is  associated with galaxies,
extinction is correlated with foreground galaxies and thus is  likely
able to produce a false quasar-galaxy or galaxy-galaxy cross correlation
signal. (2) Contrary to optical
surveys, redshifts of these galaxies can be
precisely determined by the redshifted wavelength of the 21cm
hyperfine transition line,  
with no added observational cost.  Precision measurement
of galaxy redshift is required for the precision prediction of lensing
statistics. It also allows the luxury of removing close galaxy pairs,
which is crucial to measure cosmic magnification in the presence of
the strong auto correlation
of galaxies.   (3) At high redshift 
the magnification effect is enhanced, since galaxies at higher redshifts
are more strongly lensed and have a steeper flux distribution. Combined
with the large sample of galaxies, shot noise can be overcome. (4)
The auto correlation function of cosmic magnification can be measured
to high accuracy. The prediction of the cosmic magnification auto
correlation is free of galaxy bias prior and as robust as the
prediction of cosmic shear power spectrum.

The goal of this paper is modest, which is to demonstrate the feasibility to
do precision  lensing measurement using 21cm emitting
galaxies. Methods discussed in this paper are by no mean optimal
and results presented  are conservative. Efforts toward optimizing
lensing measurement is presented in  the companion
paper\citep{Zhang05}, where we show that cosmic magnification of 21cm
emitting galaxies can not only do better than  of cosmic magnification
optical
galaxies, but also can likely do better than cosmic shear of optical
galaxies.

Throughout this paper, we adopt a flat $\Lambda$CDM universe with
$\Omega_m=0.3$, $\Omega_{\Lambda}=1-\Omega_m$, $\sigma_8=0.9$,
$h=0.7$ and initial power index $n=1$, as is consistent with WMAP
result \citep{Spergel03}. We take  Canadian Large Adaptive Reflector
(CLAR)\footnote{http://www.clar.ca/} and SKA as our targets to forecast the ability of
future radio 21cm surveys to measure cosmic magnification. The
instrumental parameters and survey patterns of these two surveys have not
been completely fixed yet. For CLAR, we adopt a system temperature $T_{\rm
  sys}=30$ K, effective collecting area $A_{\rm eff}=5\times 10^4\  {\rm
  m}^2$ and field of view $1\ {\rm deg}^2$. For SKA, we adopt the same
value of $T_{\rm sys}$ and field of view and adopt $A_{\rm
  eff}=6\times 10^5\  {\rm m}^2$.  

\section{Detecting galaxies in radio surveys}
\label{sec:nz}
Though the universe is highly ionized, there are still large amounts of
neutral gas present in galaxies. The typical HI mass is around $\sim
10^9M_{\sun}$ \citep{Zwaan97}. HI rich galaxies appear in the radio band
by emitting the 21 cm hyperfine line resulting from the transition of
atomic hydrogens from spin 1 ground state to spin zero ground
state. The spontaneous 
transition rate is $A_{21}=2.85\times 10^{-15}{\rm s}^{-1}$. 

Each emission line has negligible width. But since neutral gas has
rotational and thermal motions, the
integrated intrinsic line width is not negligible. It turns out that the
line width of  21 cm emission is mainly  
 determined by  the rotation of HI gas, which has a typical velocity
 $\sim 100 $ km/s.  The Doppler effect by thermal
 motions cause a velocity 
 width  $w\sim 
c \sqrt{k_BT/m_H}\la 10 \sqrt{T/10^4 {\rm K}}$ km/s.  The actual
emission line width is  
 determined by many parameters, such as the total mass of galaxies,
 redshift and inclination angle of the HI disk. For simplicity, we assume
 that the combined intrinsic  line width is $w=100$ km/s. The choice
 of $w$ affects the prediction of detection efficiency of HI emitting
 galaxies. But the results shown in this paper should not be changed
 significantly by more realistic choice of $w$. 

Since the velocity (frequency)
 resolution of radio surveys can be much higher than the intrinsic
 line width, the observed flux could have a nontrivial dependence on
 the bandwidth, whose modeling requires detailed description of HI
 distribution in galaxies. To
 avoid this complexity, we adopt bandwidth to be larger or equal to
 redshifted 21cm  line width, which changes from $w$ at $z$ to  $w/(1+z)$ at
 $z=0$, or in frequency space, to $\Delta \nu=w \nu_{21}/c(1+z)$,
 where $\nu_{21}=1.4$ Ghz is the 21cm frequency. The total 21 cm flux
 of HI rich galaxies is  
\ba
\label{eqn:s21}
S_{21}&=&\frac{g_2A_{21}}{g_1+g_2} \frac{N_{\rm
HI}E_{21}}{4\pi D^2_L(z)}(\frac{w}{c}\frac{\nu_{21}}{1+z})^{-1}
\\
&= & 0.023{\rm mJy}\frac{M_{\rm HI}}{10^{10}M_{\sun}}
(\frac{c/H_0}{\chi(z)})^2 \frac{100 {\rm km}/s}{w(1+z)}\ .\nonumber  
\ea
Here, $g_1=1$ and $g_2=3$ are the degeneracies of atomic hydrogen spin
$0$ and $1$ ground state. $E_{21}=h\nu_{21}$ is the energy of each 21cm
photon. $N_{\rm HI}=M_{\rm HI}/m_{\rm HI}$ is the total number of
neutral 
hydrogen of a galaxy with total hydrogen mass $M_{\rm
  HI}$. $D_L(z)=\chi(z) (1+z)$ is the luminosity distance and $\chi$
is the comoving angular distance.  $H_0=100 h$ km/s/Mpc is the
Hubble constant at present time.  

Instrumental noise scales as $\Delta
  \nu^{-1/2}$, where $\Delta \nu$ is the bandwidth. So a larger
  bandwidth helps to beat down instrumental noise. But if the bandwidth is larger than the 21 cm line width, the
  signal is diluted and scales as $1/\Delta \nu$. Thus, 
the highest signal to noise ratio is gained when the bandwidth $\Delta
\nu$ is equal to the redshifted  bandwidth of integrated 21cm emission line,
  $w \nu_{21}/c(1+z)$. The 
system noise per beam is 
\ba
S_{\rm sys}&=&\frac{\sqrt{2}T_{\rm sys}}{\eta_c A_{\rm
    eff}}\frac{k_B}{\sqrt{\Delta 
\nu t}} \\
&\simeq& 0.032{\rm mJy} \frac{T_{\rm sys}}{30 {\rm K}}\frac{5\times
  10^4
  {\rm m}^2}{A_{\rm eff}}(\frac{100{\rm km/s}}{w/(1+z)}\frac{{\rm
  hour}}{t})^{1/2} \ .\nonumber  
\ea
Here, $\eta_c$ is the correlator efficiency, which is adopted as
  $\eta_c=0.9$. 
The instrumental beam area is $A_b\sim
\lambda^2/A_{\rm eff}$. For CLAR, $A_b\simeq 10^{'2}(\nu_{21}/\nu)^2$. For SKA,
$A_b\simeq 1^{'2}(\nu_{21}/\nu)^2$.  21cm emitting regions have
typical size $\sim 30$ kpc/h or $\sim 1^{''}$ at $z\sim 1$. Thus, the
size of 21cm emitting regions is much smaller than the beam size. The
  observed flux is thus the total flux of each galaxies. In
this case, the calculation of the detection threshold is
straightforward. 

If we choose those peaks with flux above $nS_{\rm sys}$ ($n$-$\sigma$
selection threshold), the minimum HI mass selected is 
\ba
M_{\rm HI,min}&=&n\times 1.39\times 10^{10} M_{\sun}
\left(\frac{\chi(z)}{c/H_0}\right)^2 (1+z)^{3/2} \nonumber \\
&&\times  \sqrt{\frac{w}{100{\rm km/s}}\frac{\rm hour}{t}}\frac{T_{\rm
    sys}}{30 {\rm K}}\frac{5\times 
  10^4
  {\rm m}^2}{A_{\rm eff}}\ .
\ea

We assume that the HI mass function follows the Schechter function 
found at $z=0$ \citep{Zwaan97}
\be
n(M,z)dM=n_0(z)\left(\frac{M}{M_*(z)}\right)^{-\gamma}\exp\left(-\frac{M}{M_*(z)}\right)dM
\ .
\ee 
We fix $\gamma=1.2$. \citet{Zwaan97} found that
$M_*(z=0)=10^{9.55}h^{-2}M_{\sun}$ and 
$n_0(z=0)=0.014  h^{3}{\rm Mpc}^{-3}$. There is little solid
measurement of $n_0(z)$ and $M_*(z)$ other than at local
universe. But for this form of the mass function,  there exists a
tight relation 
between $\Omega_{\rm HI}$, the cosmological neutral hydrogen density
with respect to the present day critical density, and $n_0M_*$: 
\be
\label{eqn:omega_HI}
\Omega_{\rm HI}h=2.1\times
10^{-4}\frac{n_0(z)}{n_0(z=0)}\frac{M_*(z)}{M_*(z=0)}\ .
\ee

\begin{figure}
\epsfxsize=9cm
\epsffile{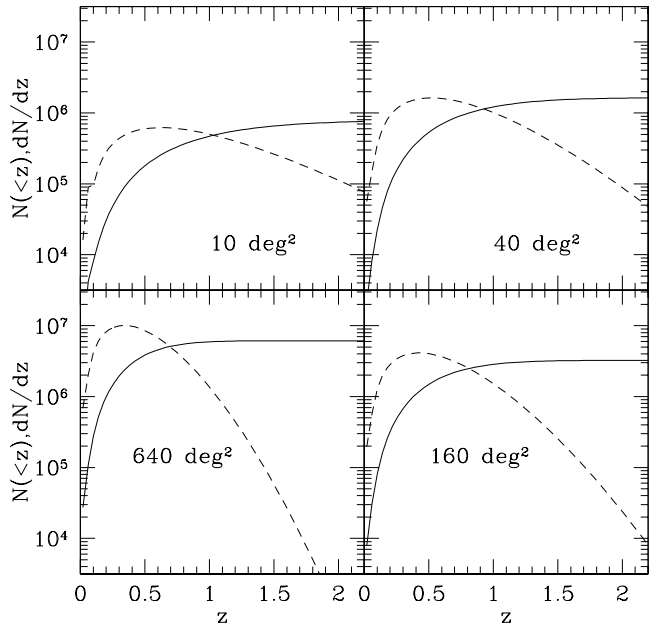}
\caption{The predicted abundance of 21cm emitting galaxies above
  $4\sigma$ detection threshold of CLAR 5 year survey.  The solid
  lines are the 
  cumulative number 
  distribution $N(<z)$
  and the dashed lines are the differential distribution $dN/dz$.  The
  total number of observed galaxies is of the order 
  $10^6$ and the number of galaxies at $z>1$ is of the order $10^5$. We assume no
  evolution in the mass function. If realistic evolution model is
  considered, the total number of galaxies at $z>1$ can increase by a
  factor of 5 or more.  \label{fig:CLAR_nz}} 
\end{figure}

Observations of
damped Lyman-$\alpha$ systems  and Lyman-$\alpha$ limit systems
measure $\Omega_{\rm HI}$ from $z=0$ to $z\sim 4$. Combining 
Eq. \ref{eqn:omega_HI} and these observations, one can put constraints
on $n_0$ and $M_*$. These observations found that  $\Omega_{\rm HI}$ increases 
by a factor of $5$ toward $z\sim 3$ and then decreases toward higher
redshift (\citet{Zwaan97, Rao00,Storrie-Lombardi00,Peroux03} and data
compilation in \citet{Peroux03, Nagamine04}). Either the increase of
$n_0$ or $M_*$ increases the 
 detectability of 21cm emitting galaxies. Thus, estimations based on
 the assumption of no evolution should be regarded as conservative
 results. We will show that even in this conservative case, future
 radio surveys such as CLAR and SKA still allow precise measurement of
 galaxy-galaxy lensing (\S \ref{sec:cross}) and in the case of SKA,
 the precise measurement of lensing power spectrum (\S \ref{sec:cross}
 \& \ref{sec:auto}) and lensing bispectrum (\S 
 \ref{sec:bispectrum}). If we adopt evolution
 models implied by and consistent with observations, even CLAR is able
 to measure the lensing power spectrum (\S \ref{sec:evolution}).

The number of galaxies detected depends on the selection threshold
 $n$. If we choose $4\sigma$ detection threshold ($n=4$), CLAR could
 detect $\sim 10^{6}$-$10^7$ galaxies in a 5-year survey. A deeper
 survey (smaller sky coverage) detects a larger fraction of high redshift
 galaxies. But since the survey volume is smaller, the total number of
 high redshift galaxies is not necessarily higher. A survey area around
 $100\ {\rm deg}^2$ is optimal to detect high $z$ galaxies. For a $160
\  {\rm deg}^2$ survey area, $\sim 10^6$ galaxies at $z> 1$ can be
 detected. SKA is about 10 times more sensitive than CLAR and can
 detect two orders of magnitude more galaxies. For a $1600\ {\rm deg}^2$
 survey area, $\sim 10^8$ galaxies at $z>1$  and $\sim 
10^7$ galaxies at $z>2$ can be detected. As a reminder, these
 estimations are extremely conservative. The number of galaxies
 detected at $z>1$ can be enhanced by a factor of $5$ or more if the evolution
 effect is considered. 

\begin{figure}
\epsfxsize=9cm
\epsffile{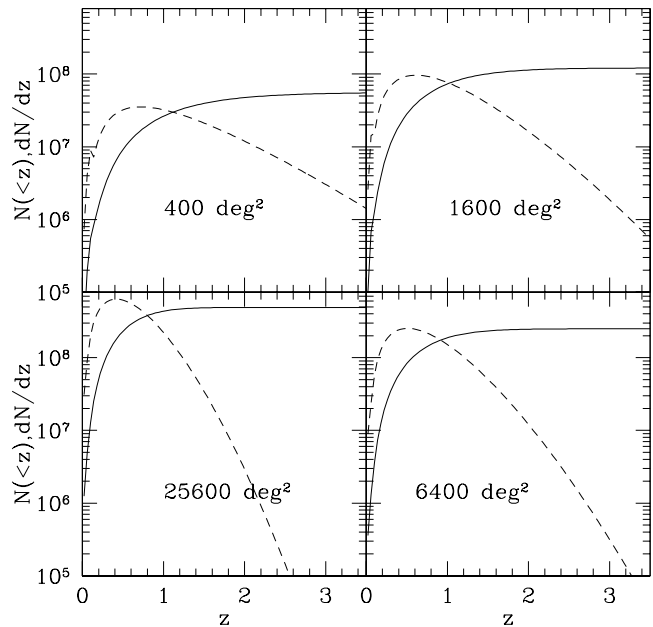}
\caption{Similar as Fig. \ref{fig:CLAR_nz}, but for SKA 5 year
  survey.  The total number of observed galaxies is of the order
  $10^8$ and  the number of galaxies at $z>1$ is of the order $10^7$.
  We assume no evolution in the mass function. If a realistic
  evolution model is 
  considered, the total number of galaxies at $z>1$ can increase by a
  factor of 5 or more. \label{fig:SKA_nz}} 
\end{figure}

Some peaks above the selection threshold are caused by noise. 
 $N_{\rm noise}$, the number of false peaks,  has strong dependence on
the detection  threshold. If we choose $n=1$( 1$\sigma$ detection), the number
of false peaks of noise is 
\be
N_{\rm noise}\sim \frac{4\pi f_{\rm sky}}{A_{\rm pixel}}
\frac{\nu}{\Delta v}  \frac{{\rm
  Erfc}(1.0/\sqrt{2})}{2}\la 10^{10} f_{\rm sky}\frac{1^{'2}}{A_{b}}\ .
\ee
 For CLAR deep surveys which
cover $f_{\rm sky}\sim 1\%$ of the sky, $N_{\rm noise}\la 
10^7$, which is still less the number of detected galaxies above
 $1\sigma$ threshold. For SKA with several
thousand square degree sky coverage,  $N_{\rm noise}$ above $1\sigma$
 is $\sim 10^8$, which is slightly less than  the total number of galaxies
above  $1\sigma$ threshold.  In CMB measurements, signal-to-noise
per pixel is often chosen to be 1, which maximizes return on the power spectrum
measurement.  This would correspond to a detection threshold 
for which the false-positive rate is $50\%$. In this sense, we
 can choose $n=1$ as our selection threshold.  The selection threshold
 problem can be dealt with in a more sophisticated way. The survey measures a
 three-dimensional map of the sky.  Each pixel 
in that map will have some significance of detection.  Clearly the large
number of low significance pixels do collectively contain information,
if they can be averaged in a meaningful way. \citet{Zhangt05}
describe one such algorithm to extract the luminosity function deep into
the noise.

With these large samples of galaxies, the galaxy clustering can be
precisely measured. This allows constraining cosmology through the
baryon oscillation \citep{Abdalla04}. The redshift distortion of
galaxies allows the measurement of galaxy velocity power spectrum to
high accuracy\footnote{Zhang\& Pen, 2005, in preparation}. In this paper we
do not attempt to utilize all information of these 
galaxies. Our primary goal is to demonstrate the feasibility of using future
21cm radio survey to do precision lensing measurement. We notice that
the properties of 
lensing magnification depends strongly on the mass threshold or equivalently
$n$. We will explore different $n\geq 1$ to find suitable choice for
different quantities. For $n\ga 3$, since the number of false peaks is
 much smaller than the number of detected galaxies, one can safely 
neglect all errors caused by false detections. But for $n\la 3$, one
 has to take them into account.

\section{Cross correlations of different redshift bins}
\label{sec:cross}
\subsection{Cosmic magnification preliminary}

Cosmic magnification changes the galaxy number overdensity $\delta_g$
to \footnote{In
this expression, we neglect all high order terms throughout this
paper. These terms increase the lensing signal
\citep{Menard03a} and thus improve the correlation measurement.}
\be
\label{eqn:magnification}
\delta^L_g=\delta_g+2(\alpha-1)\kappa+O(\kappa^2)\ .
\ee
Here, $\alpha\equiv -f^{'}(>F_c)F_c/f(F_c)$, $f(>F)$ is the number of
galaxies brighter than $F$, $F_c$ is the flux limit adopted and
$\kappa$ is the lensing convergence.   Since $\langle
\kappa\rangle=0$, to the accuracy of $\langle \kappa^2\rangle \sim
10^{-4}$, lensing does not change the averaged $f(>F)$. Thus $\alpha$
is effectively an observable.

%The lensed $f^L(>F)$ is related
%to the unlensed $f(>F)$ 
%by 
%\ba
%f^L(>F)&=&\int f(>\frac{F}{\mu}) P(\mu)d\mu\\
%&=&f(>F)+(4f^{'}F+2f^{''}F^2)\langle\kappa^2\rangle+O(\kappa^3)\ .\nonumber
%\ea
%Here, $P(\mu)$ is the probability distribution function of $\mu$. 
%Lensing can demagnify or magnify galaxies.  To the first order
%approximation, $\langle
%u\rangle=2\langle \kappa\rangle=0$ and thus  does not change the
%averaged $f(>F)$. Even including higher order corrections, since
%$\langle\kappa^2\rangle\la 10^{-3}$, 
%$f^L(>F)=f(>F)$ is still an excellent approximation.  One can further
%iterate the above expression to get a  more accurate estimate of
%$f(>F)$. So, $\alpha$ is a direct observable. Thus the prediction of
%the measured correlations is completely determined by the clustering
%of matter and galaxies. Such correlations are then powerful tool to
%constrain cosmology. 

The usual method to  eliminate the intrinsic correlation of close 
galaxy pairs
is to cross   correlate  galaxies in two separate redshift
bins. Galaxy 
peculiar velocity shifts the position of galaxies in the redshift
space by $c\Delta z \la 10^3 {\rm km/s}$. Galaxy correlation
becomes negligible at
scales $r\ga 100 h^{-1}$Mpc, which corresponds to $\Delta z\sim 0.03
[H(z)/H_0][r/100 h^{-1} {\rm Mpc}]$. Choosing $\Delta  z\ga 0.05$, residual
correlations caused by intrinsic galaxy correlation can be safely
neglected.   With accurate measurement of 21cm emitting galaxy
redshift, this can be done straightforwardly.

%\begin{figure}
%\epsfxsize=9cm
%\epsffile{SKA_alpha.eps}
%\caption{Similar as Fig. \ref{fig:CLAR_alpha}, but for
%  SKA. \label{fig:SKA_alpha}} 
%\end{figure}

The distribution of foreground galaxies traces lenses of background
galaxies which cause the magnification effect. Thus there exists a
correlation between the 
background magnification and foreground galaxies. This galaxy-galaxy
correlation is the
correlation generally considered in the literature. There exists
another correlation. Both foreground galaxies and background galaxies
are lensed by intervening matter. Thus there exists the background
magnification-foreground magnification correlation. The combined
correlation is 
\ba
\label{eqn:3DdeltaL}
\langle \delta^L_g({\bf \theta}_f,z_f)
\delta^L_g({\bf \theta}_b,z_b)\rangle&=&2(\alpha_b-1)\langle \kappa_b
\delta({\bf \theta}_f,z_f\rangle \\
&+&4(\alpha_f-1)(\alpha_b-1)\langle \kappa_f \kappa_b\rangle\ .\nonumber
\ea
Here the subscript $f$ and $b$ denote foreground and background
respectively. The 
first term in the right side of the equation is the
magnification-galaxy correlation and the second term is the
magnification-magnification correlation.

The observed galaxy surface density is
\be
\Sigma_g=\int_{z_{\rm min}}^{z_{\rm max}} n_g(z) (1+\delta^L) dz\ .
\ee
Its correlation function is Eq. \ref{eqn:3DdeltaL} weighted by the
differential galaxy number distribution $n_g$ and clustering signal.
By Limber's approximation, the
corresponding 2D angular power spectrum of the magnification-galaxy cross
correlation is given by
\ba
\label{eqn:clmg}
\frac{l^2C^{\mu g}_l}{2\pi}&=&\frac{3\Omega_m
  H^2_0}{2c^2}N_f^{-1}\frac{\pi}{l} \\
&\times&\int_{z_{f,{\rm min}}}^{z_{f,{\rm max}}}
\Delta^2_{gm}(\frac{l}{\chi_f},z_f)G_b(z_f)n_g(z_f)\chi_f dz_f\ . \nonumber
\ea
Here, $\Omega_m$ is the present day matter density with respect to the
  cosmological critical density. $N_f=\int n_g(z_f)dz_f$ is the total number of
  foreground galaxies. $\Delta^2_{gm}=b_gr_g\Delta^2_m$ is the
  galaxy-matter cross correlation power 
spectrum. We assume $b_gr_g=1$ ($b_g$ is the galaxy bias and $r_g$ is
  the  cross correlation
  coefficient between galaxies and dark matter).  The matter power
  spectrum $\Delta^2_m$ is calculated using the BBKS
  transfer function \citep{Bardeen86} and its nonlinear evolution  is calculated by Peacock-Dodds fitting formula
\citep{Peacock96}. $G_b$, the kernel of the background magnification effect, is
  given by 
\be
G_b(z)=\frac{1+z}{N_b}\int_{z_{b,{\rm min}}}^{z_{b,{\rm max}}}
  w(\chi,\chi_b)n_g(z_b)2(\alpha(z_b)-1)dz_b\ .
\ee
Here, $ w(\chi,\chi_s)$ is the lensing geometry
function. For the flat universe we adopt, 
  the geometry function is simplified to 
  $w(\chi,\chi_s)=\chi(1-\chi/\chi_s)$. The strength of the
  magnification effect relies on both 
  the strength of lensing, which prefers background galaxies with
  higher $z$, and $\alpha-1$, which prefers deeper slope of the mass
  function at the mass threshold. 

\begin{figure}
\epsfxsize=9cm
\epsffile{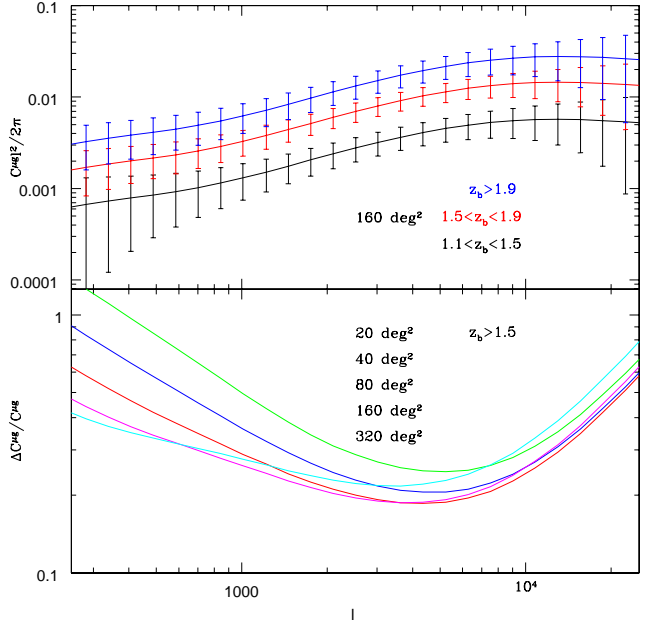}
\caption{The predicted accuracy of cosmic magnification-galaxy cross 
  correlation power spectrum $C^{\mu g}$ measured by CLAR 5 year
  survey. Foreground galaxies ($0.5<z_f<1.0$) are selected with a
  $2\sigma$ selection  threshold.
The optimal survey coverage is around several hundred square
  degrees. For such configuration, $C^{\mu g}$ can be measured to
  $\sim 20\%$ accuracy at $l$ around several thousand (lower panel,
  bin size $\Delta l=0.2l$).  $C^{\mu\mu}$ is several percent of
  $C^{\mu g}$ and does not show up in the plot. In top panel, we further
  split background 
  galaxies into several redshift bins. This lensing tomography allows the
  measure of the evolution of matter 
  distribution. It also allows  the measure of relative change of the
  comoving distance as a 
  function of $z$ to better than $10\%$ in
  three redshift bins at $z_b>1$.   \label{fig:CLARmg}}
\end{figure}

The power spectrum of the corresponding magnification-magnification cross
correlation is:
\ba
\label{eqn:clmm}
\frac{l^2C^{\mu\mu}_l}{2\pi}&=&\left(
\frac{3\Omega_mH^2_0}{2c^2}\right)^2\frac{\pi}{l}\\
&\times&\int_0^{z_{f,{\rm max}}}
\Delta^2_{m}(\frac{l}{\chi},z)G_b(z)G_f(z)\chi d\chi \ .\nonumber
\ea
Here, $G_f$, the kernel of foreground magnification effect, is given
by 
\be
G_f(z)=\frac{1+z}{N_f}\int_{z_{f,{\rm min}}}^{z_{f,{\rm max}}} w(\chi,\chi_f)n_g(z_f)2(\alpha(z_f)-1)dz_f
\ee
where $N_f$ is the total number of foreground galaxies. The amplitude of 
$C^{\mu \mu}$ relies on both the lensing signal of foreground and
background galaxies. The higher the  $z_f$ and $z_b$, the stronger the
correlation signal. It also depends strongly on $\alpha_f-1$ and
$\alpha_b-1$.  If the mass threshold is larger, $\alpha-1$ is
generally larger. 

\begin{figure}
\epsfxsize=9cm
\epsffile{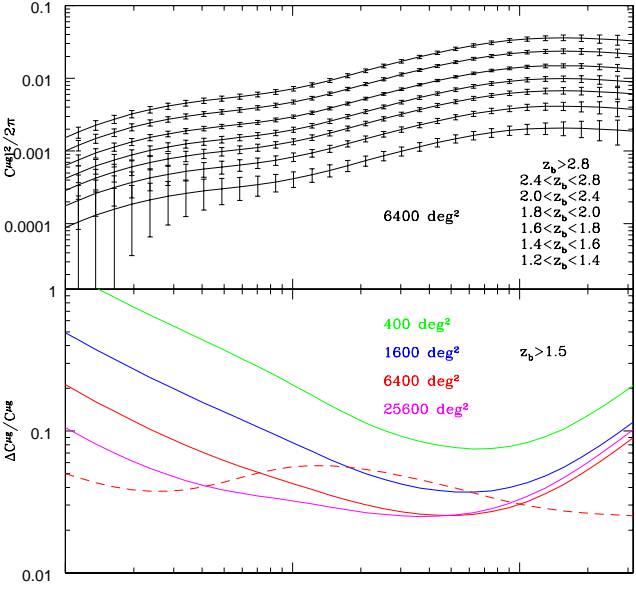}
\caption{Similar as Fig. \ref{fig:CLARmg}, but for SKA 5 year
  survey. Foreground galaxies are chosen to be 
  at $0.7<z_f<1.0$.
To measure $C^{\mu g}$, optimal sky
  coverage should be around ten thousand degrees. Under such
  configuration, $C^{\mu g}$ can be measured to several percent accuracy at
  $1000\la l\la 10^4$. The dashed line in the lower panel  is
  $C^{\mu\mu}/C^{\mu g}$ for $6400$ ${\rm deg}^2$ sky coverage.   The
  dependence  of comoving distance $\chi(z)$ on 
  redshift can be measured to be better than $1\%$ in 7
  redshift bins at $z>1$.   \label{fig:SKAmg}}
\end{figure}

 Depending on the choice of
foreground bins and  foreground galaxy selection criteria (which
determines $\alpha$ and thus the strength of magnification effect),
either $C^{\mu g}$ or $C^{\mu\mu}$  can dominate.  In \S
\ref{subsec:mg}, we discuss cases where $C^{\mu g}$ dominates and
in \S \ref{subsec:mm}, we discuss cases where $C^{\mu\mu}$
dominates.  

\subsection{Magnification-galaxy power spectrum}
\label{subsec:mg}
For a sufficiently wide foreground galaxy redshift distribution with low median
redshift, the lensing effect and the magnification ($\alpha-1$) effect
are both small. Generally, in this case, 
$C_l^{\mu\mu}$  is much smaller than $C_l^{\mu g}$. In this section, we fix
the foreground galaxy distribution ($0<z_f<1.0$) and vary the
background galaxy redshift distribution.  For this choice of
foreground galaxy distribution, $C^{\mu \mu}$ is  $\sim 1\%$ of
$C^{\mu g}$ (Fig. \ref{fig:SKAmg}). The
correlation signal peaks at $l\simeq 10^4$, where the fluctuation is
$\sim 10\%$. 

$C_l^{\mu g}$ is the projection of $\Delta^2_{gm}$ along the line of
sight (Eq. \ref{eqn:clmg}).  Given a cosmology, $\Delta^2_{gm}(k,z)$
can be extracted using the inversion methods  applied to galaxy
surveys and lensing surveys (e.g. \citet{Dodelson02,Pen03b}).
$\Delta^2_{gm}(k,z)$ contains valuable information of galaxy
clustering and can put strong constraint on halo occupation distribution.

Cosmological information is also carried in the geometry term of
lensing, in our case, $\chi$ in $C^{\mu g}$ and $C^{\mu\mu}$.  
By fixing the foreground galaxy distribution and varying the
background galaxy distribution, one 
can isolate $\chi_b$ from $\Delta^2_{gm}$ and measure the dependence of
$\chi_b$ on $z_b$. This method allows an independent and robust
constraint on cosmology \citep{Jain03,Zhang03}.

The statistical error in the $C^{\mu g}$ measurement\footnote{At frequencies below 1.4 Ghz, radio sources have smooth continuum
 spectra, so they can be subtracted away in the frequency space
 \citep{Wang05}. Their only effect is to contribute to $T_{\rm sys}$
 and cause a fluctuation in $T_{\rm sys}$ across the sky. This can
 introduce a false correlation, but it is routine in observation to
 eliminate this effect and thus we do not consider it in this paper.} is 
\ba
\label{eqn:dcmug}
\Delta C^{\mu g}_l=\sqrt{\frac{C_{\mu g}^2+(C^b_g+C^{b}_{\rm shot})(C^f_g+C^f_{\rm
    shot})}{(2l+1)\Delta l f_{\rm
      sky}}}\ .
\ea
Here, $C^b_g$ is the auto correlation power spectrum of the background
galaxies, which 
includes contributions from $\langle \delta^b_g\delta^b_g\rangle $,
$\langle \kappa_b\kappa_b\rangle$ and $\langle
\delta^b_g\kappa_b\rangle$.  For high redshift background galaxies,
$\langle \kappa_b\kappa_b\rangle$ is not negligible comparing to the
intrinsic galaxy  correlation  $\langle \delta^b_g\delta^b_g\rangle $
and thus has to be taken into account. $C^b_{\rm shot}=4\pi f_{\rm
  sky}[1+N_{\rm noise}^b/N_b]/N_b$ is the background shot noise power spectrum. The
extra factor $1+N_{\rm noise}^b/N_b$ accounts for the
contaminations of false peaks in the selected background sample. These false
peaks does not correlate with each other, so their only effect is to
increase the shot noise.  For selection thresholds above $2$-$\sigma$,
false positive rate is small and this extra factor can be
neglected. $C^b_f$ is the auto correlation power 
spectrum of the foreground  galaxies.

Statistical errors strongly depend on $f_{\rm sky}$ (when fixing total
observation time). This
dependence  is
complicated. (1) $f_{\rm sky}$ affects
$\alpha(z_f)$. Shorter integration time per unit area is required in
order to survey for a larger sky area. This increases system noise per
beam and thus increases the selection mass threshold of high z
galaxies. Since  the mass function is steeper at higher mass,
$\alpha(z_f)$ increases and $C^{\mu g}$ increases. (2) $f_{\rm sky}$
affects the relative distribution of galaxies. Larger $f_{\rm sky}$
survey detects relatively more low z galaxies. Since $C_l^{\mu g}$ is
proportional to $\Delta^2_{gm}$ 
weighted by the distribution of foreground galaxies and matter clustering is
stronger at lower $z$, larger sky coverage tends to increase $C^{\mu
  g}$. But on the other hand, the lensing effect is smaller for lower
$z$ galaxies. This has the effect to decrease $w(\chi,\chi_b)$ and
thus decrease 
$C^{\mu g}$. Furthermore, the noise terms $C^b_g$ and $C^f_g$
increase. (3) $f_{\rm sky}$ affects the cosmic variance. (4)
$f_{\rm sky}$ affects the total number of foreground and background
galaxies and thus changes the shot noise.  The lower panels of
fig. \ref{fig:CLARmg} 
  \& \ref{fig:SKAmg} show the dependence of $\Delta C^{\mu g}/\Delta
  C^{\mu g}$ on sky coverage. If the sky coverage is too small, the
  cosmic variance is large. If the sky coverage is too
  large, too few background galaxies can be detected. Shot noise
  begins to dominate at relatively large scales. 
The choice of selection threshold $n$ ($n$-$\sigma$) also affects the
statistical errors. Larger $n$ increases
$\alpha$ and thus increases 
$C^{\mu g}$. But it also decreases the number of detected galaxies and
  thus increases shot noise.

Both CLAR and SKA can measure $C^{\mu g}$ precisely (fig. \ref{fig:CLARmg} 
  \& \ref{fig:SKAmg}). For CLAR, the optimal sky coverage is around
  several hundred square degree. $C^{\mu g}$ can be measured to $\sim
  20\%$ accuracy for bin size $\Delta l=0.2l$. It can also measure
  $C^{\mu g}$ at several redshift bins and allows isolating
  geometry. For SKA, the optimal sky coverage is 
  around ten thousand square degree. $C^{\mu g}$ can be measured to
  $\sim \%$ accuracy. The size of background
  redshift bins can be as narrow as $\Delta z\la 0.1$. The change of
  $\chi_b(z)$ can be precisely measured at $z\ga 1$.

The best result from optical  galaxy surveys by far is an $8\sigma$
detection by SDSS, using an optimal quasar weighting function
\citep{Scranton05}. CLAR will detect much less 
galaxies and cover much smaller sky area, but even CLAR can reach
$\sim 5$-$\sigma$ at more than 10 independent bins of width 
$\Delta l=0.2l$. If adopted similar background galaxy weighting
function, the result can be further improved. So CLAR and SKA can do
much better than optical galaxy surveys.
This is well explained by Eq. \ref{eqn:dcmug}.
(1) The number of $z>1$ CLAR galaxies is of the order $10^6$ and is
much higher than the number of SDSS quasars ($\sim 2\times
10^4$). This significantly 
reduces the shot noise term $C^b_{\rm shot}$ in
Eq. \ref{eqn:dcmug}. (2) CLAR foreground galaxies have broad redshift
distribution, which roughly matches the lensing kernel of background
galaxies and thus amplifies the cross correlation signal. For SDSS,
quasars lie at $z\ga 1$ while most galaxies locate at $z\ll
0.5$. Since these quasars are mainly lensed by matter at $z\ga 0.5$,
the cross correlation signal is weak and $C^{\mu g}\ll
\sqrt{C^b_gC^f_g}$. This is likely the dominant  reason
why the measured cross correlations by 2dF and SDSS
\citep{Myers03,Scranton05} do not have as good S/N as one would naively
expect from their large galaxy and quasar samples.

%If
%foreground galaxies locate far away 
%from the redshift range where the dominant lensing signal of
%background galaxies comes from, these galaxies do not contribute to
%$C^{\mu g}$ and only contribute to noise $C^{f}_{g}$. To have large
%lensing signal, background galaxies must locate at high redshift $z_b\ga
%1$. The lensing signal mainly comes from $z\ga 0.5$ redshift
%range. The contribution of foreground galaxies at $z_f\la 0.5$ to cross
%correlation signal $C^{\mu g}$ is small. But these galaxies are
%strongly clustered and cause large statistical error to $C^{\mu
%  g}$. 
%In the measurement of quasar-galaxy cross correlation using
%optical surveys (e.g. \citet{Myers03,Scranton05}), quasars lie at $z\ga
%1$ while most galaxies locate at $z\ll 0.5$.  This results in  $C^{\mu g}\ll
%\sqrt{C^b_gC^f_g}$. 

It is difficult for optical surveys to detect large number of high z
galaxies. But 21cm radio surveys can. This is an inherent  advantage of
21cm radio surveys to measure cosmic magnification. Furthermore, since
21cm galaxies have precise redshift measurement, one can weight
foreground galaxies deliberatively to optimize the cross correlation
measurement. This is another inherent  advantage of
21cm radio surveys.

\begin{figure}
\epsfxsize=9cm
\epsffile{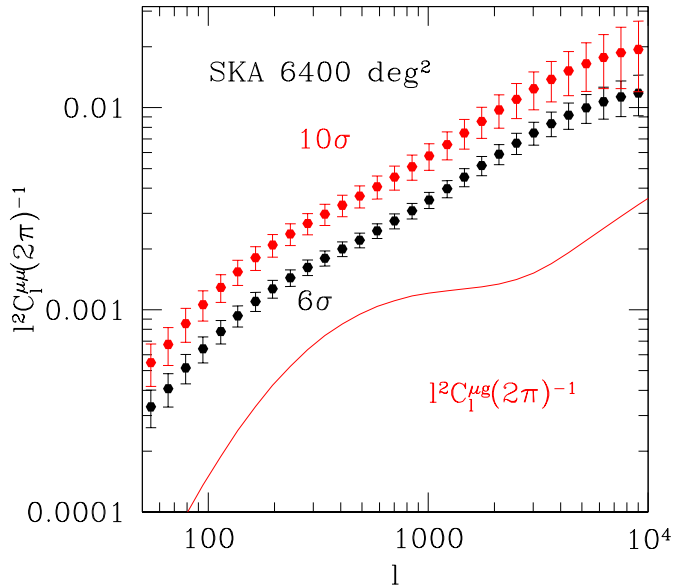}
\caption{The predicted accuracy of $C^{\mu \mu}_l$, magnification-Magnification cross correlation
  power spectrum in two redshift bins,  by SKA 5 year
  surveys. The measurement of $C^{\mu \mu}_l$ allows constraining
  cosmology and studying matter clustering without the complexity of
  galaxy bias. We choose foreground galaxies at $1.6\leq z_f\leq 2.0$ and
  background galaxies at $2.0\leq z_b\leq
  4.0$. We treat $C^{\mu g}_l$ as contaminations. Since $C^{\mu \mu}$
  depends on $\alpha-1$ of foreground galaxies while $C^{\mu g}$ does
  not, we  vary the
  selection threshold of foreground galaxies to change
  $C^{\mu\mu}/C^{\mu g}$. We fix the selection threshold of background
  galaxies as
  $4\sigma$. For $6\sigma$ and $10\sigma$ selection
  threshold of  foreground, 
  galaxies, systematic errors ($C^{\mu g}$) is comparable to
  statistical errors (error-bars of data points).  \label{fig:SKAmm}}
\end{figure}

\subsection{Magnification-Magnification power spectrum}
\label{subsec:mm}
The main strength of the cosmic shear power spectrum and bispectrum to
constrain cosmology lies in the fact that the prediction of these
quantities only relies on the matter clustering whose theoretical
understanding is robust.  The prediction of $C^{\mu\mu}$ is as
straightforward as the prediction 
of the cosmic shear power spectrum, which does not require the
knowledge of complicated galaxy bias, as $C^{\mu g}$
does. So, the measurement of $C^{\mu\mu}$ would allow robust constraints
on cosmological parameters and dark matter clustering, as cosmic shear
power spectrum does.  In this section, we will show that SKA is
straightforward to measure the 
magnification (foreground)-magnification (background) cross
correlation power spectrum $C^{\mu\mu}$. 

 For the purpose of constraining
cosmology using $C^{\mu \mu}$, $C^{\mu g}$ should be treated as
contamination and marginalized over. $C^{\mu\mu}_l$
depends on the magnification strength of foreground galaxies, while
$C^{\mu g}_l$ does not. By  increasing  the redshifts of
foreground galaxies, the lensing signal increases and $\alpha-1$ also
increases, due to higher mass selection threshold at higher redshift,
thus $C^{\mu\mu}_l$ increases with respect to $C^{\mu g}_l$. Since
$C^{\mu g}$ is proportional to the strength of matter clustering, it
decreases when increasing redshifts of foreground galaxies. But these
requirements to increase $C^{\mu\mu}$ with respect to $C^{\mu g}$ can
be  at odds
with the requirement to reduce statistical errors. For example,
increasing the mass selection threshold or redshifts of foreground
galaxies  reduces $N_f$ and thus increases shot noise.

We try different foreground redshift bins, selection
threshold and sky coverage to optimize the measurement of $C^{\mu
  \mu}$ such that  $C^{\mu g}/C^{\mu \mu}$ is smaller or comparable to
the statistical error. For SKA, it is indeed possible to measure
$C^{\mu\mu}$ and control both systematic errors ($C^{\mu g}$) and
statistical errors to $10\%$ level (Fig. \ref{fig:SKAmm}). Since such
measurement requires large number of high z 
galaxies, it is extremely difficult for optical surveys to realize it. 

\begin{figure}
\epsfxsize=8cm
\epsffile{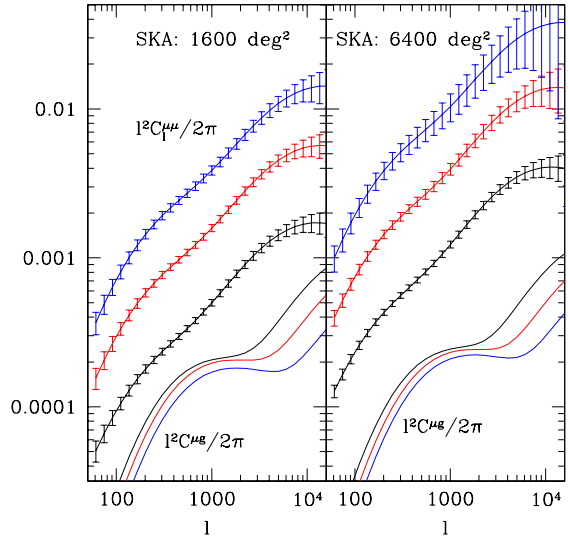}
\caption{Auto correlation angular power spectrum of galaxies. We
  disregard close pairs within redshift separation $\Delta z<0.1$ and thus eliminate
  intrinsic galaxy clustering.  We try different galaxy distribution. Lines with error bars, from bottom to top, correspond to $z>1.5$,
  $z>2.0$, and $z>2.5$, respectively. For our choice of $4\sigma$ selection
  threshold and $z>1.5$, $C^{\mu\mu}$ dominates over $C^{\mu g}$.
For higher redshift, the luminosity
  function is steeper at the limiting flux. Larger $\alpha-1$ then
  increases $C^{\mu\mu}$ with respect to $C^{\mu g}$. On the other
  hand, high redshift galaxies are mainly lensed by low z matter
  distribution. The higher the redshift, the less likely that galaxies
  can be lensed by matter distribution in the same redshift bins. This
  also increase $C^{\mu\mu}$ with respect to $C^{\mu g}$. We assume no
  evolution in the HI mass function. Realistic evolution scenarios
  would result in more galaxies and thus allow better
  measurement.\label{fig:SKA_auto}}
\end{figure}

\section{Auto correlation}
\label{sec:auto}
With the precision measurement of galaxy redshift and large amount of
high z galaxies, one can extract the cosmic magnification from the
galaxy auto correlation measurement. In the correlation estimator, we
throw away pairs with redshift separation $|z_1-z_2|\leq \Delta 
z_c$. We choose $\Delta z_c=0.1$, which corresponds to comoving
separation $r_c\simeq 180 h^{-1}$Mpc at $z=1$ and $r_c\simeq 100
h^{-1}$Mpc at $z=2$. In Fourier space, this
corresponds to cut off the power at $k_c\la 1/r_c\la 
0.01h/$Mpc. Applying the Limber's equation in Fourier space, the angular
fluctuation at multipole $l$ is contributed by the spacial  fluctuation
at $k=l/\chi$. Then  an effective cut off $k_c$ in
Fourier space corresponds to an effective cutoff at $l_c=k_c\chi\la
20$. So, under the Limber's approximation, one can
neglect the  residual intrinsic correlation of galaxies at $l\ga 20$.

One can further quantify the residual intrinsic clustering. The angular 
 correlation it produces is
\ba
w^c_{\rm IC}(\theta)&=&2 \int_{z_{{\rm min}}}^{z_{{\rm max}}}
 n^2_g(z)\frac{dz}{d\chi/dz}\\
&&\times  \int_{r_c}^{\infty}
 \xi_g(\sqrt{\chi^2\theta^2+(\Delta \chi)^2},z)d\Delta \chi \ .\nonumber
\ea
Here, $\xi_g$ is the galaxy correlation function. Numerical
 calculation shows that $|w^c_{\rm IC}(\theta)|$ is smaller than 
 $10^{-5}$ at all scales.  For example, $|w^c_{\rm IC}(\sim 1^{'})|$ is less than $\sim
 10^{-5}$ and $|w^c_{\rm IC}(\sim 1^{\circ})|$ is less than ${\rm
 several}\times 
 10^{-7}$. One can further convert $w_{\rm IC}^c(\theta)$ to the
 corresponding $C_l$. We find that 
 $|l^2C_l/(2\pi)|\la 10^{-5}$ at all scales. Specifically,  
 at $l\sim 100$, $|l^2C_l/(2\pi)|\la {\rm several} \times 10^{-6}$ and at
 $l\sim 1000$, $|l^2C_l/(2\pi)|\la {\rm several} \times 10^{-6}$. The
 angular fluctuation caused by residual galaxy intrinsic clustering is
 roughly $1\%$ of $C^{\mu\mu}$ at $l\sim 100$ and much less than $1\%$
 at smaller scales. So, the close pair removal procedure  effectively
 eliminates all intrinsic galaxy clustering.  

The auto correlation function is composed of two parts, the one arising
from the auto correlation of cosmic magnification and the one from the cross
correlation between cosmic magnification  and $\delta_g$. The
magnification auto correlation power spectrum is
\be
\frac{l^2C^{\mu\mu}_l}{2\pi}=\left(
\frac{3\Omega_m H^2_0}{2c^2}\right)^2\frac{\pi}{l}\int_0^{z_{{\rm max}}}
\Delta^2_{m}(\frac{l}{\chi},z)G(z)^2f_2^2(z)\chi d\chi 
\ee
where $N$ is the total number of galaxies in the corresponding
redshift bin. This expression differs from Eq. \ref{eqn:clmm} only by a factor
$f_2^2(z)$, which arises from the close pair removal. $f_2(z)$ is
given by 
\ba
f^2_2(z)&=&[\frac{(1+z)}{NG(z)}]^2\int
4w(\chi,\chi_1)(\alpha_1-1)w(\chi,\chi_2)(\alpha_2-1)\nonumber\\
&\times & n_g(z_1)n_g(z_2)\Theta(|z_1-z_2|-\Delta z_c) dz_1dz_2  \ .
\ea
The function $\Theta(x)=0$ if $x<0$ and $\Theta(x)=1$ if $x>0$. $G(z)$
is defined analogous to $G_b$ and $G_f$.

The power spectrum of the magnification-galaxy cross correlation function
is 
\ba
\frac{l^2C^{\mu g}_l}{2\pi}&=&\frac{3\Omega_m
  H^2_0}{2c^2}N^{-1}\frac{\pi}{l} \int_{z_{{\rm min}}}^{z_{{\rm max}}}\\
&\times&
\Delta^2_{gm}(\frac{l}{\chi(z)},z)G(z)f_1(z)n_g(z)\chi(z) dz\ . \nonumber
\ea
The effect of close pair removal is carried by $f_1(z)$ 
\ba
f_1(z)&=&\frac{1+z}{NG(z)}\int_{z_{{\rm min}}}^{z_{{\rm max}}}
  w(\chi,\chi_b)\\
&\times& n_g(z_b)(\alpha(z_b)-1)\Theta(|z_b-z|-\Delta z_c)dz_b\ . \nonumber
\ea

Again, as explained in \ref{subsec:mm}, for the purpose of
constraining cosmology, we treat $C^{\mu g}$ as contaminations of
$C^{\mu\mu}$. For the SKA,  both the statistical errors of $C^{\mu\mu}$
and systematics errors ($C^{\mu g}$) can be controlled to better than $10\%$
level, if we only use  $z\ga 1.0$ galaxies
(Fig. \ref{fig:SKA_auto}). For CLAR, the detectability of $C^{\mu\mu}$
is sensitive to the 
evolution of HI mass function. Assuming the conservative no evolution
model,  $C^{\mu 
\mu}$ can be detected at the several $\sigma$
level(fig. \ref{fig:CLAR_evolution}). Though this close pair removal
method is successful for 21cm emitting galaxies, it is essentially
infeasible for optical galaxies because of (1) insufficient number of
high z optical galaxies and (2) inaccurate photo-z redshifts and too
expensive spectroscopic redshifts.

\section{Cosmic magnification bispectrum}
\label{sec:bispectrum}
Lensing bispectrum contains
valuable and often complimentary information on cosmology and the
large scale structure, comparing to the 2-point correlation power spectrum
\citep{Bernardeau97,Hui99,Bernardeau02,Menard03c,Pen03,Takada04}.
But current data only allows low significance detection  of the skewness at
several angular scales \citep{Bernardeau02,Pen03,Jarvis04}. 21cm radio
surveys can do much better. In this
section, we will show its feasibility by focusing  on the bispectrum
of galaxies in the same redshift bins. We throw away close 
  pairs with $|z_i-z_j|<0.1$ where $z_i$ ($i=1,2,3$) is the redshift
  of each galaxies.

The bispectrum comes  from four parts,  $\langle \mu\mu\mu\rangle$, $\langle \mu\mu
g\rangle$, $\langle \mu g g\rangle$  and $\langle ggg\rangle$.
The  $\langle \mu g g\rangle$ and $\langle ggg\rangle$ terms are 
negligible following similar argument  in \S \ref{sec:auto}. The  $\langle \mu\mu\mu\rangle$ term contributes a bispectrum 
\ba
B_{\mu\mu\mu}(l_1,l_2,l_3)&=&\left(\frac{3\Omega_mH_0^2}{2c^2}\right)^3\int_0^{z_{\rm max}}
G^3(z)f^3_3(z)\chi^{-4}\nonumber \\
&&\times 
B_{\delta}(\frac{l_1}{\chi},\frac{l_2}{\chi},\frac{l_3}{\chi};\chi)
d\chi \ .
\ea
Here, $f_3(z)$ takes the effect of close pair removal into
account. $B_{\delta}(k_1,k_2,k_3;\chi)$ is the matter density
bispectrum, which is  calculated adopting the
fitting formula of \citet{Scoccimarro01}. 

The $\langle \mu\mu g\rangle$ term contributes another bispectrum 
\ba
B_{\mu\mu g}(l_1,l_2,l_3)&=&3\left(\frac{3\Omega_mH_0^2}{2c^2}\right)^2\int_{z_{\rm
    min}}^{z_{\rm max}}
G^2(z)f_4^2(z)\chi^{-4} \nonumber\\
&& \times
b_g B_{\delta}(\frac{l_1}{\chi},\frac{l_2}{\chi},\frac{l_3}{\chi};\chi)
\frac{n_g(z)}{N}dz\ .
\ea
Here, $f_4(z)$ takes the effect of close pair removal into account. We
explicitly show the galaxy bias $b_g$ in the above equation, though we
adopt $b_g=1$ in the estimation. The factor $3$ comes from the
permutation of $\langle \mu_i \mu_jg_k\rangle$.

The shot noise of the bispectrum is
\be
B_{\rm shot}(l_1,l_2,l_3)=C_N^2+C_N(C_1+C_2+C_3)\ .
\ee
Here, $C_N=4\pi f_{\rm sky}/N$ is the shot noise power spectrum. Full
evaluation  of the sample variance of bispectrum involves integrating 6-point 
nonlinear density correlation function. Since we  do not have  robust
theoretical predictions or simulation results of 6-pt nonlinear correlation
function, we only consider the Gaussian sample variance.  The statistical
error of corresponding bispectrum $B_{123}$ is
\be
\Delta B_{123}=\sqrt{\frac{2}{N_{123}}\left(C_1C_2C_3+B_{\rm
    shot}^2\right)}\ .
\ee
Here, $C_i=C^{\mu\mu}(l_i)+C^{\mu g}(l_i)$. $N_{123}$ is the number of
independent combination of 
$l_1,l_2,l_3$ used to obtain the averaged $B_{123}$.
For a rectangle
survey area with $x$ axis size $\theta_x$ and $y$ axis size
$\theta_y$, independent modes are ${\bf l}=(2\pi m/\theta_x, 2\pi
n/\theta_y)$, where $m,n=0, \pm 1, \pm 2, \cdots$. There is a constraint that $l_3=-l_1-l_2$, then the total
number of independent combination is
\ba
N_{123}&=&dm_{1x}dn_{1x}dm_{2x}dn_{2y}\\
&&=\left(\frac{\theta_x}{2\pi}\frac{\theta_y}{2\pi}\right)^2
l_1dl_1l_2dl_2 2\pi d\theta_{12}\nonumber
\ea
where $\theta_{12}$ is the angle between ${\bf l}_1$ and ${\bf
  l}_2$. We only show the result of ${\bf l}_1\simeq {\bf l}_2\simeq
{\bf l}_3$, for which, we choose $\Delta l=0.2 l$ and $\Delta
\theta_{12}=\pi/18$ ($10^{\circ}$). 

\begin{figure}
\epsfxsize=9cm
\epsffile{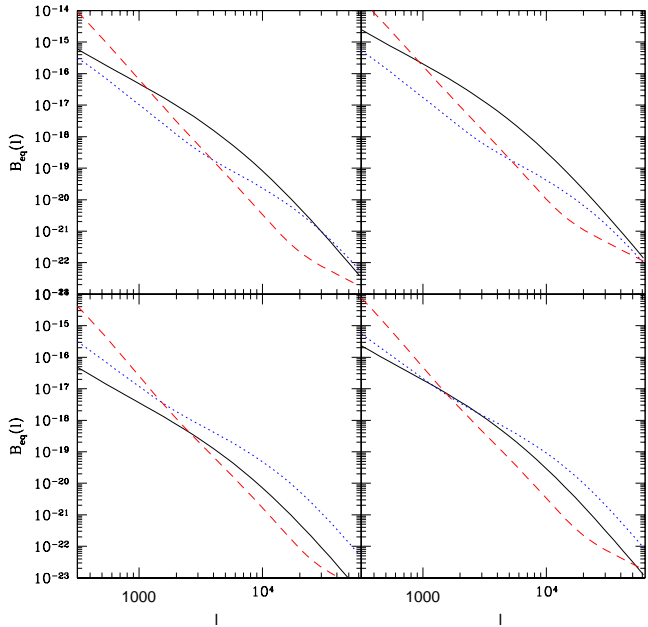}
\caption{Forecasted measurement of bispectrum by SKA. We assume no
  evolution in the HI mass function.   The solid lines are
  $B_{\mu\mu\mu}$ of equilateral configuration ($l_1=l_2=l_3$), the
  dot lines  are  $B_{\mu\mu g}$ and the
  dash lines are the statistical error. For statistical error, we adopt
  the $l$ bin 
  size $\Delta l=0.2 l$ and the angular bin size $10^{\circ}$. We try
  different galaxy redshift ranges and selection thresholds. The
top left, top right, bottom left and bottom right panels are the
  results of $(3\sigma, 
  z>1.5)$, $(5\sigma, z>1.5)$,  $(3\sigma, z>1.0)$ and  $(5\sigma,
  z>1.0)$, respectively. \label{fig:B}} 
\end{figure}

At large scales ($l\la 1000$), cosmic variance prohibits the
measurement of $B_{\mu\mu\mu}$ and $B_{\mu\mu g}$. But at scales
$l\sim 10^4$, either $B_{\mu\mu\mu}$ or  $B_{\mu\mu g}$ or both can be
measured to be better than $10\%$ accuracy by SKA
(Fig. \ref{fig:B}). As expected, higher z and/or higher selection
threshold result in higher lensing signal and amplify $B_{\mu\mu\mu}$
with respect to $B_{\mu\mu g}$.  The computation of  three point
function may appear computationally 
challenging, requiring the enumeration of $N^3$ triangles with $N\sim 10^8$.
Recently, linear algorithms have been devised which resolve these
problems \citep{Zhangl03}. Again the requirements of precise
redshift measurement and high z galaxies prohibit the measurement of
lensing bispectrum through cosmic magnification of optical galaxies.

\section{Evolution Effect}
\label{sec:evolution}
We have demonstrated the feasibility of measuring lensing power
spectra and bispectrum  in cross correlation of galaxies in two
redshift bins and 
in auto correlation of galaxies in the same redshift bin. We  caution
on that these results (fig. \ref{fig:CLAR_nz}, \ref{fig:SKA_nz},
\ref{fig:CLARmg},\ref{fig:SKAmg},\ref{fig:SKAmm} \&
\ref{fig:SKA_auto},\ref{fig:B}) should be regarded as
conservative estimation of the power of radio surveys to measure
cosmic magnification. There are several reasons. One is that in this
paper we only use galaxies in 
certain redshift ranges and above certain selection threshold. Better
measurement of $C^{\mu \mu}$ and $C^{\mu g}$ can be obtained by cross
correlating galaxies above $1\sigma$ detection threshold at all
redshifts. To estimate how much can one gain requires the design of
careful weighting on different redshifts and luminosity (HI
mass). This work is beyond the scope of this paper. Another reason is
that we have assumed no evolution in the HI mass function. Evolution
effect is very likely to improve the accuracy of lensing measurement
by providing many more detected galaxies. We will investigate the
evolution effect in this section.
\begin{figure}
\epsfxsize=9cm
\epsffile{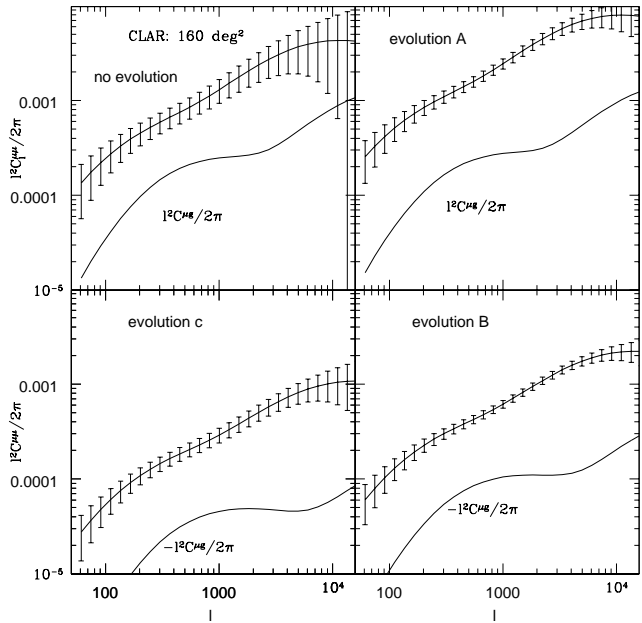}
\caption{$C^{\mu\mu}$ and $C^{\mu g}$  that would be measured in the
  same redshift 
  bin by CLAR. We try different evolution models, as explained in \S \ref{sec:evolution}. Top left panel
  assumes no evolution and we choose galaxies selected above $2\sigma$
  at $z>1.5$. Other panels assume evolution models explained in the
  text.  Top right panel uses galaxies above $3\sigma$ at
  $z>1.5$. Bottom right panel uses galaxies above $2\sigma$ at
  $z>2.0$,  Bottom left panel uses galaxies above $1\sigma$
  at$z>2.5$. \label{fig:CLAR_evolution}}
\end{figure}

As discussed in \S \ref{sec:nz},  the observed $\Omega_{\rm HI}h$
shows a factor of 5 increase from $z=0$ to $z\sim 3$.  Its evolution
can be approximated as $g(z)=(1+z)^{2.9}\exp(-z/1.3)$. Thus we have a
constraint of  $n_0(z)
M_*(z)=n_0(z=0) M_*(z=0)  
g(z)$.   There is little solid constraint on  the evolution of $n_0$ or
$M_*$ separately. But the observation of damped Lyman-$\alpha$ systems
and Lyman-Limit systems provides some indirect constraints. Damped
Lyman-$\alpha$ systems have HI column density $N_{\rm HI}\geq 2\times
10^{20}\ 
{\rm cm}^{-2}$. If the size of the corresponding HI regions is $\sim
30\ 
{\rm kpc}/h$, then the total HI mass is $\sim 10^{10} M_{\sun}$. Thus these
damped Lyman-$\alpha$ systems are likely part of corresponding massive
HI (proto-)galaxies. On the other hand, Lyman-Limit systems have much
smaller HI column density and are likely part of less massive HI
galaxies. The ratio of damped Lyman-$\alpha$ systems abundance with
respect to Lyman-limit systems decreases after $z=3$. This implies
that there may be fewer massive HI galaxies after $z\sim 3$ and thus
an evolution of $M_*(z)$.

Since the constraint to either $n_0$ or $M_*$ is weak and it is likely
that  both $n_0(z)$ and $M_*(z)$ evolve, we explore three 
evolution scenarios. (A) No evolution in
$M_*(z)$. $n_0(z)=n_0(z=0)g(z)$.  (B) No  evolution in $n_0(z)$.
$M_*(z)=M_*(z=0)g(z)$.
(C). $n_0(z)/n_0(z=0)=M_*(z)/M_*(z=0)=g(z)^{1/2}$.  

The number of $z>1$ galaxies increases by at least a factor of
5 for these evolution scenarios. Taken the evolution effect into
account, even CLAR can measure 
$C^{\mu\mu}$ to $\sim 10\%$ accuracy (systematic and statistical,
fig. \ref{fig:CLAR_evolution}).

\section{Discussion}
\label{sec:discussion}
We further address that the results shown in this paper only
utilize a small fraction of cosmic magnification information contained
in 21cm emitting galaxy distribution. (1) We
only tried several bins of 
galaxy redshift distribution and several galaxy selection threshold to
demonstrate that cosmic magnification can be measured to high
accuracy. To utilize the full lensing information, one needs to divide
galaxies into many redshift bins and selection threshold bins. One
then measures (N-point) auto correlation functions of each bins
and cross correlation functions between different bins. In principle,
one can develop optimal weighting scheme to combine all measurements
to get the best measurement of lensing statistics and lensing-galaxy
statistics. (2) We did not attempt to separate the cosmic magnification auto
correlations and the cosmic magnification-galaxy correlations. These
two classes of correlation have different dependence on the selection
threshold of galaxies. These dependences are straightforward to
predict and can be applied to separate two components.  Such
component separation improves the robustness of constraining cosmology
and large scale structure significantly. The cosmic
magnification auto  correlations and the geometry of cosmic
magnification-galaxy correlations \citep{Jain03,Zhang03} are ideal to
constrain cosmology and matter clustering. The amplitude and  angular scale
dependence  cosmic
magnification-galaxy correlations are ideal to constrain halo
occupation distribution. Advanced
analysis methods are required to address the above two issues and to
utilize full information of cosmic magnification information in 21cm
emitting galaxy distribution. (3) We note that individual galaxies can
be resolved with SKA, this allows the measurement of cosmic
shear. It may also allow an independent measure  of cosmic magnification.  SKA resolution allows 
the measurement of an inclination angle.  If galaxies at high redshift
also follow a Tully-Fisher relation, the lensing effect can also be
large compared to shot noise and can be extracted.  Since we do not
know the evolution of the 
scatter in the Tully-Fisher at high redshift, we do not use this information
in this paper.  It is likely that real surveys can do significantly better
than our estimates.

Utilizing all information of cosmic magnification information  in 21cm
emitting galaxy distribution, the relative error  of cosmic
magnification will be much smaller than what shown in this paper. Upon
this precision era, one needs to improve the theoretical prediction to
better than $1\%$ accuracy and at the same time, understand possible $1\%$
systematics.  

Our prediction of cosmic magnification is simplified in two ways. (1)
We only considered the leading order
term of cosmic magnification (Eq. \ref{eqn:magnification}). Higher
order terms are known to be capable of generating $10\%$ effect
(e.g. \citet{Menard03a}) and have to be included to interpret
data at the forecast accuracy. (2) For the lensing convergence, we neglected high order
corrections caused by  lens-lens coupling and deviation from Born's
approximation. These high order corrections are known to have several
percent effect \citep{Schneider98,Vale03,Dodelson05} and should be taken into
account. 

Source-lens coupling \citep{Bernardeau98,Hamana01} can have
non-negligible effect to cosmic shear. It arises from the fact that measured cosmic
shear is always weighted by the number of observed galaxies, which also trace 
the matter distribution. But this effect does not exist in cosmic
magnification where we directly correlate the numbers of
observed galaxies at two different redshifts and directions. 

Several other approximations in our cosmic magnification measurement
only introduce negligible corrections. (1) we have assumed that the
luminosity function $f(>F)$ is 
the same everywhere and thus $\alpha$ is the same everywhere. This
picture is over simplified. $f(>F)$ can have environmental
dependence. Thus in principle $\langle \alpha\alpha\rangle\neq \langle
\alpha\rangle^2$. This affects the prediction of correlations where
two or more cosmic magnification terms present (e.g. in $C^{\mu\mu}$,
$B_{\mu\mu\mu}$ and $B_{\mu\mu g}$). But this effect is very
small. since $\alpha$ only depends on local environment, for two
redshift bins with modest separation $\Delta z\ga 0.05$, $\langle
\alpha_1\alpha_2\rangle\simeq \langle 
\alpha_1\rangle\langle \alpha_2\rangle$. The close pair removal
procedure further guarantees that even for the same redshift bin,  $\langle
\alpha_1\alpha_2\rangle\simeq \langle 
\alpha_1\rangle\langle \alpha_2\rangle$. Thus, one can safely neglect
this effect.  (2)  Residual intrinsic clustering causes  $\sim 1\%$ correction
at $l\sim 100$ (\S \ref{sec:auto})\footnote{But it causes $\ll 1\%$ correction at smaller
  scales.}. Since cosmic variance
at $l\sim 100$ is $\ga 0.01 f_{\rm sky}^{-1/2}$. One needs to worry
about this effect only for full sky surveys. Furthermore,  it can be
reduced by extrapolating  
galaxy correlation function $\xi_g$ at smaller scales measured from
the same survey  to relevant scales ($\ga 100h^{-1}$Mpc).

\section{Conclusions}

We have made simple forecasts for future radio surveys to measure
gravitational lensing.  We found that radio surveys can be precise sources
for lensing measurements, and that lensing magnification is measurable
because redshifts are known and many galaxies can be detected.  CLAR
and SKA are expected to measure the 
dark matter power spectrum and galaxy-matter cross correlation to high
accuracy.  The estimates are conservative, since evolution in HI mass
function, as suggested by observations, will improve the above results
significantly. Also, many effects will 
increase the sensitivity.  Unfortunately, these effects, which include
Tully-Fisher relations, are difficult to quantify at
high redshifts, for which we neglect in this paper.

More complete statistical information is available at lower signal to noise
levels and using non-Gaussian statistics.  We have made estimates of
the three point statistics, which appear promising, and are expected
to improve the information that can be gained by lensing.

{\it Acknowledgments}
We thank Ron Ekers and Simon Johnston for explanation of radio
experiments. We thank Scott Dodelson for helpful discussion.   
P.J. Zhang was supported by the DOE and the NASA grant NAG 5-10842 at
Fermilab.

\end{document}